\documentclass[12pt]{article}
\usepackage{epsfig,amssymb}

\newcommand{\bq}{\begin{equation}}
\newcommand{\eq}{\end{equation}}
\newcommand{\bqa}{\begin{eqnarray}}
\newcommand{\eqa}{\end{eqnarray}}
\newcommand{\ben}{\begin{enumerate}}
\newcommand{\een}{\end{enumerate}}
\newcommand{\bc}{\begin{center}}
\newcommand{\ec}{\end{center}}
\newcommand{\bqb}{\begin{eqnarray*}}
\newcommand{\eqb}{\end{eqnarray*}}

\def\lsim{\lesssim}

\def\pr#1#2#3{ Phys. Rev. ${\bf{#1}}$ (#2) #3}
\def\prl#1#2#3{ Phys. Rev. Lett. ${\bf{#1}}$ (#2) #3}
\def\pl#1#2#3{ Phys. Lett. ${\bf{#1}}$ (#2) #3}

\def\rmp#1#2#3{ Rev. Mod. Phys. ${\bf{#1}}$ (#2) #3}
\def\np#1#2#3{ Nucl. Phys. ${\bf{#1}}$ (#2) #3}
\def\zp#1#2#3{ Z. f. Phys. ${\bf{#1}}$ (#2) #3}
\def\ijmp#1#2#3{ Int. J. Mod. Phys. ${\bf{#1}}$ (#2) #3}

\def\eg{{\it e.g.\/}}

\def\etal{{\it et.al.\/}}

\global\nulldelimiterspace = 0pt

\def\wtil#1{\widetilde{#1}}

\def\swd{s^2_W}

\def\mwd{M_W^2}
\def\mw{M_W}
\def\mz{M_Z}

\def\mv{m_V}
\def\mh{m_H}
\def\mvd{m_V^2}

\def\lamQCD{\Lambda_{QCD}}
\def\t{\hat t}
\def\s{\hat s}
\def\u{\hat u}

\begin{document}
\pagenumbering{arabic}
\thispagestyle{empty}
\def\thefootnote{\fnsymbol{footnote}}
\setcounter{footnote}{1}

\begin{flushright}
PM/98-20 \\ CPT-98/P3677  \\ THES-TP 98/06 \\
hep-ph/9807563 \\
July 1998
 \end{flushright}
\vspace{2cm}
\begin{center}
{\Large\bf Glue constraining asymmetries in $W$, $\gamma$ or
$Z$ production at CERN LHC}\footnote{Partially supported by the EC contract
CHRX-CT94-0579, and by the NATO grant CRG 971470.}
 \vspace{1.5cm}  \\
{\large P. Chiappetta$^a$, G.J. Gounaris$^b$, J. Layssac$^c$
and F.M. Renard$^c$}\\
\vspace{0.7cm}
$^a$Centre de Physique Th\'{e}orique, UPR 7061,
CNRS Luminy, F-13288 Marseille,
France \\
\vspace{0.2cm}
$^b$Department of Theoretical Physics, University of Thessaloniki,\\
Gr-54006, Thessaloniki, Greece.\\
\vspace{0.2cm}
$^c$Physique
Math\'{e}matique et Th\'{e}orique,
UMR 5825\\
Universit\'{e} Montpellier II,
 F-34095 Montpellier Cedex 5.\\

\vspace*{1cm}

{\bf Abstract}
\end{center}

We propose a class of forward-backward asymmetries with respect to
the subprocess c.m. scattering angle in $V$+jet  production at
hadron colliders, with $V$ being any of
$(W,~Z,~\gamma, ~H$), which are directly proportional to
the gluon distribution $g(x)$. The informations that these
asymmetries can provide are complementary of those reachable
from measurements of the
transverse momentum and/or the rapidity distributions
of a $V$ in kinematical regimes where gluon scattering
subprocesses dominate. The accuracy which can be
reached in the $W$, $Z$ and  $\gamma $ cases,
at the upgraded Tevatron and at LHC, should allow a considerable
improvement of our
knowledge of the gluon distribution function, especially at large $x$.\\

PACS:  12.38.Bx, 14.70.Dj,  24.85.+p, 12.38.Qk \\

\def\thefootnote{\arabic{footnote}}
\setcounter{footnote}{0}
\clearpage

\section{Introduction}

One difficulty  in our effort to test the Standard
Model (SM) at a hadronic collider, and search for
any new physics (NP) beyond it,
is due to the uncertainties
pertaining to the gluon distribution \cite{MRST,
CTEQ4M, Reya98}. Remembering for example,
that $gg \to H$ provides the dominant contribution to H-production at
LHC in the region of $100~ GeV \lsim \mh \lsim 800~ GeV$
\cite{Gunion-Higgs, Dawson},
we infer  that a good knowledge of the gluon distribution is
necessary for studying the properties of the Higgs particle and
precisely estimating the backgrounds. \par

The main constraints to the gluon distribution inside a nucleon
at present, arise from DIS measurements which probe the low x
range \cite{HERA},  and at higher x from measurements  of the $p_T$
distribution of a
prompt photon produced in $pp \to \gamma X$ at $\sqrt s = 23~
GeV$ WA70 \cite{WA70}, and  in $p Be \to \gamma X$ at
$p_{lab}=530~GeV$ \cite{E706}, and also from heavy flavor
production \cite{HF}. The combinations of subprocess cross
sections contributing to prompt photon production  are
always forward-backward symmetrical in the subprocess c.m.,
and consist of
\bq
\frac{d \hat \sigma ( q \bar q \to \gamma g)}{d \hat t }\ \ ,
\label{qbarqGamma}
\eq
and
\bq
\frac{d \hat \sigma ( g q(\bar q) \to \gamma q (\bar q))}{ d \hat t}
~~+ ~~\frac{d \hat \sigma ( q(\bar q) g \to \gamma q (\bar
q))}{d \hat t}
\  \ , \label{gqGamma+}
\eq
as well as of subprocess cross sections in which the photon
comes from a quark or gluon fragmentation.
The sensitivity of such measurements on $g(x)$,
arises from the fact that   for $pp$ or proton-Nucleus
scattering, the second subprocess (\ref{gqGamma+}) dominates
over most of the $p_T$ region \cite{MRST},
provided that the produced photon
is constrained to be sufficiently isolated \cite{Gordon},
in order to reduce the magnitude of the fragmentation contribution. \par

Further information on $g(x)$,
from such forward-backward symmetrical subprocesses,  could
arise at LHC and the upgraded Tevatron. It has been shown
\cite{DPZ} that a detailed measurement of the rapidity distributions
of an inclusively produced gauge boson $W,~Z$ or $\gamma$,
accompanied by a hard jet might provide an accurate
determination of the gluon structure function. In principle, the same
kind of information could also arise from Higgs+jet production,
but the expected statistics is very limited in this case.
Since the theoretical treatment of any ($V$+jet)
production for ($V~=~ W,~ Z, ~ \gamma , ~H$) is very
similar, we consider below all these cases together. \par

Thus, if in a  ($V$+jet) pair production, we restrict ourselves to
measurements of the $p_T$ and/or  rapidity distributions of $V$,
then we are only sensitive to the
combinations of subprocess cross sections\footnote{For
the moment we disregard contribution from
photon  fragmentation, to which we come back below.}
\bq
\frac{d \hat \sigma ( q \bar q \to V g)}{d \hat t} \ \ ,
\label{qbarqV}
\eq
and
\bq
\frac{d \hat \sigma ( g q(\bar q) \to V q (\bar q))}{d \hat t}
~~ + ~~ \frac{d \hat \sigma ( q(\bar q) g \to V q (\bar q))}{d \hat t}
\  \ , \label{gqV+}
\eq
while for Higgs production we also get contributions from
\bq
\frac{d \hat \sigma ( g g \to H  g)}{d \hat t} \ \ .
\label{ggHg}
\eq
These combinations of the subprocess cross sections
lead to ($V$+jet) distributions which are  symmetrical with respect
to the ($V$+jet) c.m. rapidity $\bar y$, as well as
with respect to $\cos \theta^*$, where $\theta^*$
is the subprocess c.m. scattering angle. \par

We want to stress in this paper that in addition to these
quantities, there exist a
contribution proportional to the combination
\bq
\frac{d \hat \sigma ( g q(\bar q) \to V q (\bar q))}{d \hat t}
~~ - ~~ \frac{d \hat \sigma ( q(\bar q) g \to V q (\bar q))}{d
\hat t}
\  \ , \label{gqV-}
\eq
of the subprocess cross sections, which induces ($V$+jet)
distributions which are {\it antisymmetric} with respect to $\bar y$, as
well as with respect to $\cos \theta^*$.
It turns out that these distributions   are directly sensitive
to $g(x)$ (in fact they are directly proportional to it)
and supply independent additional information, especially
in the large x range, which to our knowledge has not
been used so far, certainly due to the lack of statistics
in present colliders. Of course, they also provide new
information for the other parton distributions.
The aim of the present paper is to study these quantities.
To achieve this goal we construct for each ($V$+jet) final state, a
forward-backward asymmetry with respect to $\cos \theta^*$,
which is thus a function of the subprocess energy squared
$\s \equiv M^2 $, as well as an odd function of the
rapidity $\bar y$ of the center of mass.   \par

In calculating the aforementioned asymmetries for the $V=W,~Z,~H$ cases,
our philosophy is to  keep the leading QCD contribution for the ($V$+jet)
final state, and  include the antenna pattern effect arising
from the additional ($V$+jet+ soft gluon) final state
integrated in a suitable phase space region \cite{QCD, Khoze},
as a rough estimate of higher order QCD corrections. It turns
out that the exact  size of this antenna phase space region,
is  not very important for the considered asymmetries.
For  the ($\gamma $+jet) case, we  also include the photon
fragmentation contribution, whose effect is again not very
important, provided that a sufficiently strong isolation cut is
imposed on the produced photon. Such an isolation is anyway
desired, in order to increase the sensitivity of the above
asymmetry on the gluon distribution. \par

The content of the paper is the following. In Section 2, the
formalism is presented containing the definition of the
forward-backward asymmetry for the $V$+jet production, and including also
discussions of the possible antenna pattern effects and the
photon fragmentation contribution. In Sect. 3 the sensitivity
of the above asymmetry to the gluon distribution is discussed,
while the conclusions are given in Sect. 4.  Finally
the subprocess cross sections and quark distributions for the various
$V$ cases, are given in the Appendix.\par

\section{Formalism}
\subsection{The  $V$+jet contribution.}

The generic subprocess contributing to the $pp \to V~ jet~...$
($V=W,~Z,~\gamma, ~H$) cross section (apart from the photon
fragmentation case discussed separately), is written as
\bq
a(p_1)~ + ~ b(p_2) ~ \to ~ V(p_3) ~+ ~c(p_4) \ \ ,
\label{abVc}
\eq
where the momenta are indicated in parentheses and the masses of
the partons $(a,~b,~c)$ are neglected. As usual
$\hat s =(p_1+p_2)^2$, $\hat t =(p_3 - p_1)^2$, $\hat u =(p_3 -
p_2)^2$.  We also define $\tau =\hat s/ s$, where $s$ is the LHC
(or Tevatron) c.m. energy-squared.
The rapidity $\bar y$ of the c.m. of the ($V$+jet) subprocess
determines the  momentum fractions of
the incoming partons through
\bq
x_a= \sqrt {\tau} e^{\bar y} \ \ \ , \ \ \
x_b= \sqrt {\tau} e^{-\bar y} \ , \label{xaxb}
\eq
while the scattering angle in the c.m. of the same subprocess
satisfies\footnote{For a collection
of relevant  kinematical formulae see \eg\@  Appendix A3 in
\cite{HiggsLHC1}.}
\bq
   \cos \theta^* ~= ~ \frac{ \hat u -\hat t}{ \hat u +\hat t}
\ \ . \label{costheta*}
\eq  \par

The ($V$+jet) production cross section in $pp$ collisions is given by
\bqa
&&\frac{d \sigma (pp \to V~ jet)}{d\tau d\bar y d \cos \theta^*}
 = \frac{\hat s -\mvd}{2} \Bigg \{ g(x_a, Q^2) g(x_b, Q^2)
\frac{d \hat \sigma (gg \to Vg)}{d\hat t}
\nonumber \\
&& +\wtil{\Sigma}_V(x_a, x_b)
\frac{d \hat \sigma (q \bar{q}^\prime \to V g)}{d\hat t}   \nonumber \\
&& + \frac{1}{2}\left [g(x_a, Q^2) \Sigma_V(x_b)+\Sigma_V(x_a)
g(x_b, Q^2)
\right ] \left [ \frac{d \hat \sigma (g q \to V q)}{d\hat t} +
\frac{d \hat \sigma (q g \to V q)}{d\hat t} \right ]
\nonumber \\
&& + \frac{1}{2}\left [g(x_a, Q^2) \Sigma_V(x_b)-\Sigma_V(x_a) g(x_b,Q^2)
\right ] \left [ \frac{d \hat \sigma (g q \to V q)}{d\hat t} -
\frac{d \hat \sigma (q g \to V q)}{d\hat t} \right ]
\Bigg \} \ , \label{ppVjet}
\eqa
where (\ref{xaxb}, \ref{costheta*})
should be  used. In (\ref{ppVjet}),
$Q \simeq p_T/2$ is the usually preferred factorization
scale of the distribution functions which mimic the next to
leading order corrections.
The $\Sigma_V , \wtil{\Sigma}_V $
terms describe combinations of quark (antiquark) distributions
weighted by the quark electromagnetic or weak  charge,
while $d \hat \sigma$  denote the correspondingly  normalized
subprocess cross sections, for the various
$V=H,~ W,~Z,~\gamma $ cases.  These are given in the Appendix
for a $pp$ Collider, while for the $p \bar p$ Tevatron they
should be modified in an obvious way.\par

The last two terms of eq.(\ref{ppVjet}) have been constructed by
combining the corresponding gluon$\times$quark 
and the quark$\times$gluon
contributions in a symmetrical and an antisymmetrical parts.
It is  important to note that the last term in
(\ref{ppVjet}) is {\it antisymmetric} with respect to
$\bar y$ or $\cos \theta^*$ for the Tevatron or LHC Colliders,
while all other terms are symmetric in both these variables.
This last term will therefore be washed
out if we integrate over $\bar y$, as it always happens whenever
we only look at the rapidity and/or $p_T$ distribution of a
single $V$ or jet.
It will also be washed out, in case we cannot discriminate $V$
from the accompanying jet, and therefore also in dijet production.
Particularly for the dijet case, we note that so  
long we cannot discriminate between a 
gluon and a quark jet, the $\cos \theta^*$-antisymmetric term 
in (\ref{ppVjet}) vanishes identically, and 
only the symmetric part of  (\ref{ppVjet}) contributes. 
For studying  therefore the last term in (\ref{ppVjet}), which is
the main purpose of the present paper, the two high $p_T$
objects in the final state should be distinguishable,
as \eg\@  in the case
a $W,~ Z$ or $~\gamma $ production, accompanied by a high $p_T$ jet.\par

We next turn to the  corrections to the various terms in
(\ref{ppVjet}),
arising from soft gluon emission in the antenna approximation,
and from photon fragmentation in case $V=\gamma$. \par

\subsection{Adding the antenna pattern contribution.}
The hadronic antenna patterns in $V= ~ W,~Z,~\gamma $ + jet
production have been shown to provide a valuable diagnostic tool
for probing the nature of the underlying parton subprocess.
They could also provide a good tool for distinguishing between
conventional QCD and new physics production \cite{Stirling}
(large $E_T$ jet events in hadronic collisions, or
large anomalous $Q^2$ events at HERA).
Here the philosophy is different: in order to make our treatment
more realistic, we  add to the leading order calculation of
the $V$+jet cross section,  the
contribution from the soft gluon emission. Thus, for each
subprocess like in (\ref{abVc}), we consider  the
corresponding subprocess
\bq
a(p_1)~ + ~ b(p_2) ~ \to ~ V(p_3) ~+ ~c(p_4)~+~g(k) \ \ .
\label{abVcg}
\eq
The soft gluon emission in such subprocesses is controlled
by the basic antenna pattern distribution \cite{QCD}
\bq
[ij]~=~\frac{p_i \cdot p_j}{p_i\cdot k ~p_j \cdot k} \ \ ,
\label{antenna}
\eq
which should be adequate for the description of
gluonic minijets with energies
$k_0$ much smaller than those of the hard jets
\cite{Khoze}.\par

The contribution from  the ($V$+jet + soft gluon)
production cross section in $pp$ collisions is then
given by (compare (\ref{ppVjet}))
\bqa
&&k_0~ \frac{d \sigma (pp \to V~ jet)}{d\tau d\bar y d \cos
\theta^* d^3k}
 = \frac{3 \alpha_s (\hat s -\mvd)}{8\pi^2 }
 ~ \cdot  \nonumber \\
&& \Bigg \{ g(x_a, Q^2) g(x_b, Q^2) \left ([12]+[14]+[24] \right )
\frac{d \hat \sigma (gg \to Vg)}{d\hat t} \nonumber \\
& +&  \wtil{\Sigma}_V(x_a, x_b) \left( [14]+[24]-\frac{[12]}{9}
\right )
\frac{d \hat \sigma (q \bar{q}^\prime \to V g)}{d\hat t}   \nonumber \\
&+ & \frac{1}{2}\left [g(x_a, Q^2) \Sigma_V(x_b)+\Sigma_V(x_a)
g(x_b, Q^2)
\right ] \cdot \nonumber \\
&~& \Bigg [ \left ([12]+[14]-\frac{[24]}{9} \right )
 \frac{d \hat \sigma (g q \to V q)}{d\hat t} +
 \left ([12]+[24]-\frac{[14]}{9} \right )
\frac{d \hat \sigma (q g \to V q)}{d\hat t} \Bigg ]
\nonumber \\
&+& \frac{1}{2}\left [g(x_a, Q^2) \Sigma_V(x_b)-\Sigma_V(x_a)
g(x_b, Q^2)
\right ] \cdot \nonumber \\
&~& \Bigg [ \left ([12]+[14]-\frac{[24]}{9} \right )
\frac{d \hat \sigma (g q \to V q)}{d\hat t} -
 \left ([12]+[24]-\frac{[14]}{9} \right )
\frac{d \hat \sigma (q g \to V q)}{d\hat t} \Bigg ]
\Bigg \} \ . \label{ppVjetg}
\eqa  \par

Neglecting all parton masses, except $\mv$, we use the
notation $p_1^\mu=\sqrt{\hat s}/2(1,0,0,1)$,
$p_2^\mu=\sqrt{\hat s}/2(1,0,0,-1)$ for the description of the
momenta of the incoming partons $a,b$ in their c.m.
(compare (\ref{abVcg})), and the notation
\bqa
p_V^\mu & = &p_3^\mu = (E_{VT}\cosh y_V^*,~ p_T,~ 0,~
E_{VT}\sinh y_V^*) \ , \label{pV}\\
p_4^\mu &=& p_T (\cosh \eta_c^*,~ -1,~ 0, ~\sinh \eta_c^*) \ ,
\label{p4}\\
k^\mu &=& k_T (\cosh (\eta_c^*+\delta \eta),~ \cos(\delta
\varphi),~ \sin(\delta \varphi),~ \sinh
(\eta_c^* +\delta \eta)) \ , \label{k}
\eqa
for the final state particles in the  same frame \cite{Khoze}.
Here
$E_{VT}=\sqrt{\mv^2+p_T^2}$ gives the transverse energy of $V$,
and $(y_V^*, ~\eta_c^*)$ denote the
rapidities of $(V ,~ c)$ in the $(a,b)$-c.m. frame.
The final gluon
is taken to be soft, which means
\bq
 \lamQCD \ll k_0 \ll \sqrt{\hat s}/2, ~
E_{VT} \cosh y_V^*, ~ p_T\cosh \eta_c^*  \ .
\eq
Note from
(\ref{p4}, \ref{k}), that to the extent that $\delta \eta ~,~
\delta \varphi$ are small, the rapidity and azimuthal angle of
the minijet generated by the soft gluon are close to those of
the hard jet generated by parton $c$;
compare (\ref{abVcg}). \par

In terms of these variables the antenna pattern coefficients in
(\ref{antenna}) may be expressed as
\bqa
\protect [12] &= & \frac{2}{k_T^2}  \ , \label{ant12} \\
\protect [14] & = & \frac{e^{\delta \eta}}{k_T^2\left [ \cosh(\delta \eta)+
\cos(\delta \varphi) \right ] } \ ,   \ , \label{ant14} \\
\protect [24] & = & \frac{e^{-\delta \eta}}{k_T^2\left [ \cosh(\delta \eta)+
\cos(\delta \varphi) \right ] } \ , \label{ant24}
\eqa
while the phase space of the soft gluon
is determined by
\bq
\frac{d^3k}{k_0}~=~ k_T dk_T~ d \delta \eta ~d \delta \varphi \ .
\eq \par

Defining $\delta R \equiv \sqrt{\delta \eta^2+\delta \varphi^2}$,
we  choose  to integrate the antenna
activity contained in  (\ref{ppVjetg}) in the region
\bqa
\delta R_1  \leq & \delta R & \leq \delta R_2 \ , \label{R1R2} \\
k_{Tmin} \leq & k_T & \leq k_{Tmax} \  . \label{kTmaxmin}
\eqa
Adding then (\ref{ppVjet}) and the result of integrating (\ref{ppVjetg})
we get
\bqa
&&\frac{d \sigma (pp \to V~ jet)}{d\tau d\bar y d \cos \theta^*}
 = \frac{\hat s -\mvd}{2} \Bigg \{ g(x_a, Q^2) g(x_b, Q^2) I_{gg} ~
\frac{d \hat \sigma (gg \to Vg)}{d\hat t}
\nonumber \\
& + & \wtil{\Sigma}_V(x_a, x_b) I_{q\bar q}~
\frac{d \hat \sigma (q \bar{q}^\prime \to V g)}{d\hat t}   \nonumber \\
& +& \frac{1}{2}\left [g(x_a, Q^2) \Sigma_V(x_b)+\Sigma_V(x_a)
g(x_b, Q^2)
\right ] I_{gq} \left [ \frac{d \hat \sigma (g q \to V q)}{d\hat t} +
\frac{d \hat \sigma (q g \to V q)}{d\hat t} \right ]
\nonumber \\
& + & \frac{1}{2}\left [g(x_a, Q^2) \Sigma_V(x_b)-\Sigma_V(x_a)
g(x_b, Q^2)
\right ] I_{gq} \left [ \frac{d \hat \sigma (g q \to V q)}{d\hat t} -
\frac{d \hat \sigma (q g \to V q)}{d\hat t} \right ]
\Bigg \} \ , \label{ppVjet+g}
\eqa
where the  antenna contribution is contained in the parameters
\bqa
I_{gg} &=& 1+\frac{3 \alpha_s}{2 \pi^2}
\ln \left (\frac{k_{Tmax}}{k_{Tmin}}\right )
\left [ \pi (\delta R_2^2 -\delta R_1^2) +
\xi (\delta R_2,~ \delta R_1) \right ] \ , \label{Igg} \\
I_{gq} &=& 1+\frac{3 \alpha_s}{2 \pi^2}
\ln \left (\frac{k_{Tmax}}{k_{Tmin}}\right )
\left [ \pi (\delta R_2^2 -\delta R_1^2) +
\frac{4}{9}\, \xi (\delta R_2,~ \delta R_1) \right ] \ ,
\label{Igq}\\
I_{q\bar q} &=& 1+\frac{3 \alpha_s}{2 \pi^2}
\ln \left (\frac{k_{Tmax}}{k_{Tmin}}\right )
\left [-\frac{\pi}{9}\,  (\delta R_2^2 -\delta R_1^2) +
\xi (\delta R_2,~ \delta R_1) \right ] \ , \label{Iqbarq}
\eqa
expressed in terms of the function\footnote{Notice that this result
depends crucially on the fact that the antenna integration
region (\ref{R1R2}, \ref{kTmaxmin}) is chosen so that $\delta
\varphi$ in (\ref{k}) remains small, so that $k^\mu$ never goes
so close to $p_4^\mu$, to make the perturbative
treatment unreliable; compare (\ref{k}, \ref{p4}).}
\bq
\xi (\delta R_2,~ \delta R_1) =\int_{\delta R_1}^{\delta
R_2} d \delta R ~ \delta R \int_0^{2\pi}d \phi ~
\frac{\cosh (\delta R \cos \phi)}{\cosh (\delta R \cos \phi)+
\cos (\delta R \sin \phi)} \ .
\eq\par
A general comment is in order here. The quantitative predictions
on the magnitude of the colour flow are based on the hypothesis
of Local Parton Hadron Duality \cite{ADKT}, which assumes that
colour coherence effects survive the hadronization stage;  an
hypothesis well confirmed by existing data \cite{KO}.

\subsection{Photon fragmentation contribution.}

The result (\ref{ppVjet+g}) should be adequate for the  $V= W,~Z,~H$
cases. For the $\gamma$+jet case though, we should add the
contribution from processes where the final state photon comes
from the fragmentation of a quark or gluon.
Although such
processes are formally of higher order in QCD, they may
occasionally turn out to be quite important,
since the  logarithmic growth of
the fragmentation function due to scaling violations,
compensates  one power of $\alpha_s$ \cite{QCD}.
The contribution from such fragmentation to the $\gamma$+jet
cross section  arises from each  partonic subprocess
\bq
a(p_1)~ + ~ b(p_2) ~ \to ~ c(p_3) ~+ ~d(p_4) \ \ ,
\label{abcd}
\eq
in which the parton $c$  fragments  subsequently to a photon,
through a fragmentation described
by the function $D_c^\gamma (z, Q_f^2)$, where $z$ denotes the
fraction of the $c$-momentum carried by the photon.
Following \cite{Gordon} we write the
next to leading QCD order calculation of the fragmentation
contribution  to $ pp \to \gamma ~\mbox{jet}$ as
\bqa
&&\frac{d \sigma^{frag} (pp \to \gamma ~ jet)}{d\tau d\bar y d \cos
\theta^*} = \nonumber \\
&& \frac{\s}{2} \sum_{abcd}
 f_{a/p}(x_a, Q^2) f_{b/p}(x_b, Q^2) \cdot
\frac{d \hat \sigma (ab \to cd)}{d \t} \cdot
\int_{z_{min}}^1 dz D_c^\gamma (z, Q_f^2) \ ,
\label{sigmafrag}
\eqa
where $f_{a/p}(x_a, Q^2)$ denotes the $a$-parton distribution
function in a proton, at a scale $Q \simeq p_T/2$.
 $\sum_{abcd}$ refers to the summation over
the full list of subprocesses,
$q(\bar q)+q(\bar q) \to q(\bar q)+q(\bar q)$, $q(\bar q)+g\to
q(\bar q)+g$, $q\bar q\to gg$, $gg\to q\bar q$.
The expressions for the  subprocess cross sections can be found
in \cite{ES, QCD}. \par

In (\ref{sigmafrag}), $Q_f$ denotes the scale of the
fragmentation function which, according to the next to leading
order calculation in  \cite{Gordon},  is determined by
\bq
Q_f~= ~ \frac{p_T^\gamma R}{\cosh \eta_\gamma} ~
\ , \label{Qf}
\eq
where $\eta_\gamma$  is the photon rapidity  in the laboratory
frame, and $p_T^\gamma$ its transverse momentum.
In (\ref{Qf}),
$R=\sqrt{(\delta \eta)^2+(\delta \varphi)^2}$ denotes the size
of the isolation cone around the photon produced from the
$c$-fragmentation, within which the hadronic energy $E_{had}$
is constrained to be smaller than some $\max (E_{had})$,
which
in turn determines also $z_{min}$ in (\ref{sigmafrag}) through
\bq
z_{min} ~=~ 1-~ \frac{\max (E_{had}) }{E_c} \label{zmin} \ \ .
\eq

The smaller $\max (E_{had})$ is chosen, the more isolated the
photon becomes, which in turn means a smaller photon fragmentation
contribution to the cross sections. In the calculations below we
use $R=0.7$ and $\max (E_{had}) \simeq 4~GeV$
\cite{GV} in (\ref{Qf}, \ref{sigmafrag}),
which means constraining the photon to be quite isolated.
Using then standard leading order expressions
for the various partonic cross sections in (\ref{sigmafrag})
\cite{QCD}, and the photon fragmentation fit \cite{Owens}
\bqa
z D_q^\gamma (z, Q_f^2) &=& \frac{\alpha}{2\pi } \Bigg [ e_q^2
~ \frac{2.21-1.28 z +1.29 z^2}{1-1.63 \ln(1-z)}~z^{0.049}
\nonumber \\
&& +0.002 (1-z)^2 z^{-1.54} \Bigg ] \ln (Q_f^2/\lamQCD^2)
\ , \\
z D_g^\gamma (z, Q_f^2) &=&\frac{\alpha}{2\pi}~
0.0243 (1-z)^{1.03} z^{-0.97} \ln (Q_f^2/\lamQCD^2) \
\eqa
(with $\lamQCD^2=0.04 GeV^2$), we calculate the photon
fragmentation contribution to the cross
sections, which should be added to the cross section in
(\ref{ppVjet+g}). It turns out that for all numerical applications
presented below, $d\sigma^{frag}(pp \to \gamma ~jet) $  from
(\ref{sigmafrag}) is always less than 10\%
of $d\sigma (pp \to \gamma ~ jet)$ from (\ref{ppVjet+g}).\par

We are aware that our estimate of the fragmentation contribution
is very approximative. Nevertheless there are two basic reasons to
be convinced that we are not far from a precise next to leading order
calculation. Firstly, concerning photon fragmentation functions,
it has been shown in \cite{BFG} that, in the large z domain we
probe ($z \simeq 1$), the photon fragmentation functions beyond
leading order do not strongly deviate from the leading log
parametrization we have used. Secondly, with respect to the choice of
the scale of the fragmentation function and its relation to the
isolation cone, we should admit that a fully satisfactory
treatment is lacking at present \cite{PILON}. In the present
work we simply followed the results of \cite{Gordon}.

\subsection{The Asymmetry}

The forward-backward asymmetry is defined as
\bq
A^V(\tau, \bar y)~=~
\frac{\int_0^{1-\epsilon} d\cos \theta^*
\Bigg  [\frac{d \sigma (pp \to V~ jet)}{d\tau d\bar y d \cos
\theta^*}\Bigg \vert_{\theta^*}
~- ~\frac{d \sigma (pp \to V~ jet)}{d\tau d\bar y d \cos
\theta^*}\Bigg \vert_{\pi-\theta^*} \Bigg ] }
{\int_0^{1-\epsilon} d \cos \theta^*
\Bigg  [\frac{d \sigma (pp \to V~ jet)}{d\tau d\bar y d \cos
\theta^*}\Bigg \vert_{\theta^*} ~ + ~
\frac{d \sigma (pp \to V~ jet)}{d\tau d\bar y d \cos
\theta^*}\Bigg \vert_{\pi-\theta^*} \Bigg ] } \ ,
\label{Asym1}
\eq
where $\epsilon$ is a small positive number serving to exclude
the angular region around the
beam direction, where the cross sections are not measurable, and
the pertubative treatment not applicable. The $V$ production
cross section appearing in (\ref{Asym1}) is  given in (\ref{ppVjet}),
and can be refined by including 
in it the gluon bremsstrahlung contribution
at the antenna approximation as indicated  in (\ref{ppVjet+g}),
as well as the fragmentation contribution from (\ref{sigmafrag})
for the photon case. \par

As it can be seen from (\ref{ppVjet+g}, \ref{sigmafrag}),
the asymmetry $A^V(\tau , \bar y)$ is always an antisymmetric
function of $\bar y$. For the $V=W,~Z,~H$ cases, where
(\ref{ppVjet+g}) contributes (containing the leading order and the
soft gluon contributions inside the integrals $I_i$), 
it can be expressed as
\bqa
&& A^V(\tau, \bar y)~=~ \nonumber \\[0.5cm]
&& \frac{ \left [g(x_a) \Sigma_V(x_b)-\Sigma_V(x_a) g(x_b)
\right ] I_{gq} J_{gq}^{V-}}
{ \left [g(x_a) \Sigma_V(x_b)+\Sigma_V(x_a) g(x_b)
\right ] I_{gq} J_{gq}^{V+} +
 \wtil{\Sigma}_V(x_a, x_b) I_{q\bar q} J_{q\bar q}^V
+ g(x_a) g(x_b) I_{gg} J_{gg}^V} \  . \label{Asym2}
\eqa
The antenna pattern parameters  appearing here, are given in
(\ref{Igg}-\ref{Iqbarq}), while those from the various
subprocess cross sections are written as
\bqa
J_{gq}^{V-}  & = & \frac{1}{2}
\int_0^{1-\epsilon} d \cos\theta^*
\left [ \frac{d \hat \sigma (g q \to V q)}{d\hat t} -
\frac{d \hat \sigma (q g \to V q)}{d\hat t} \right ]
\ , \label{Jgq-} \\[0.2cm]
J_{gq}^{V+} & = & \frac{1}{2}
\int_0^{1-\epsilon} d \cos\theta^*
\left [ \frac{d \hat \sigma (g q \to V q)}{d\hat t} +
\frac{d \hat \sigma (q g \to V q)}{d\hat t} \right ] \ ,
\label{Jgq+} \\[0.2cm]
J_{q\bar q}^V  & = &
\int_0^{1-\epsilon} d \cos\theta^*
\left [ \frac{d \hat \sigma (q \bar{q}^\prime \to V g)}{d\hat t}
\right ] \ , \label{Jqbarq} \\[0.2cm]
J_{gg}^V  &= &
\int_0^{1-\epsilon} d \cos\theta^*
\left [\frac{d \hat \sigma (gg \to Vg)}{d\hat t}
\right ] \ , \label{Jgg}
\eqa
where of course (\ref{Jgg}) is relevant only in the Higgs case.
The $\bar y$-antisymmetry of $A^V$ is then obvious on the basis
of (\ref{xaxb}, \ref{Asym2}). Its direct dependence on
the gluon distribution $g(x)$ is also
clear  in (\ref{Asym2}).\par

 It is possible to write a
corresponding formula for the $A^\gamma $ case, but this time
the result is more complicated, since the fragmentation contribution
involves many subprocesses. \par

\section{Sensitivity to the gluon distribution.}

In principle the above asymmetries   supply
additional independent information on the parton distributions,
and should be used together with the other usual measurements to
constrain these distributions.
Since these asymmetries are proportional  to the gluon
distribution, we would like to explore here their  usefulness
for constraining $g(x)$, assuming that the quark distributions
have already been precisely determined by other means.
Thus here, we are interested in using
the above asymmetries for the various
production processes in order to improve our knowledge on the
gluon distribution $g(x)$, particularly in the large $x$ region,
where most of the uncertainties lie. We proceed as follows.  \par

\begin{figure}[p]
\vspace*{-6cm}
\[
\epsfig{file=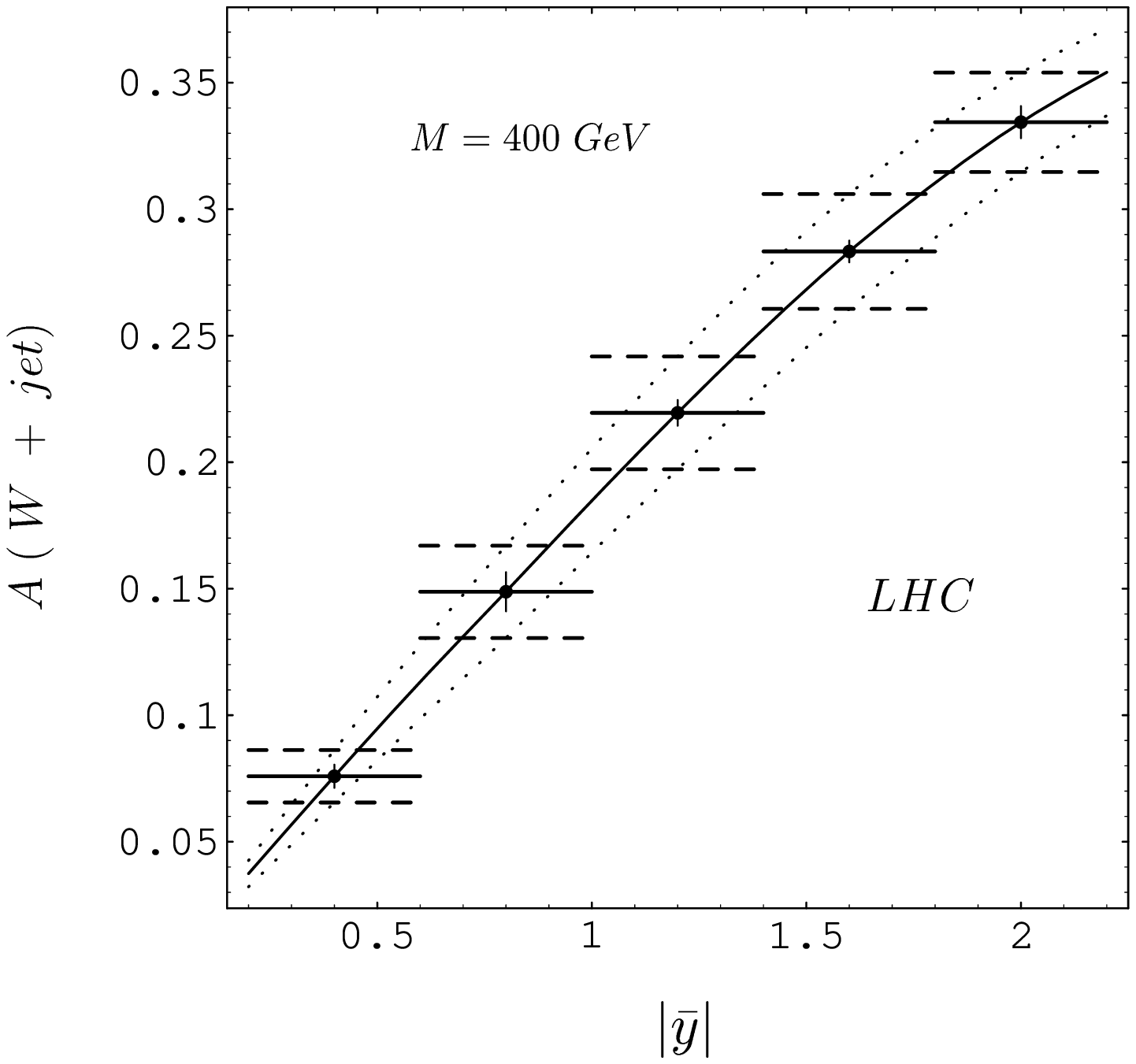,height=7.2cm}\hspace{0.5cm}
\epsfig{file=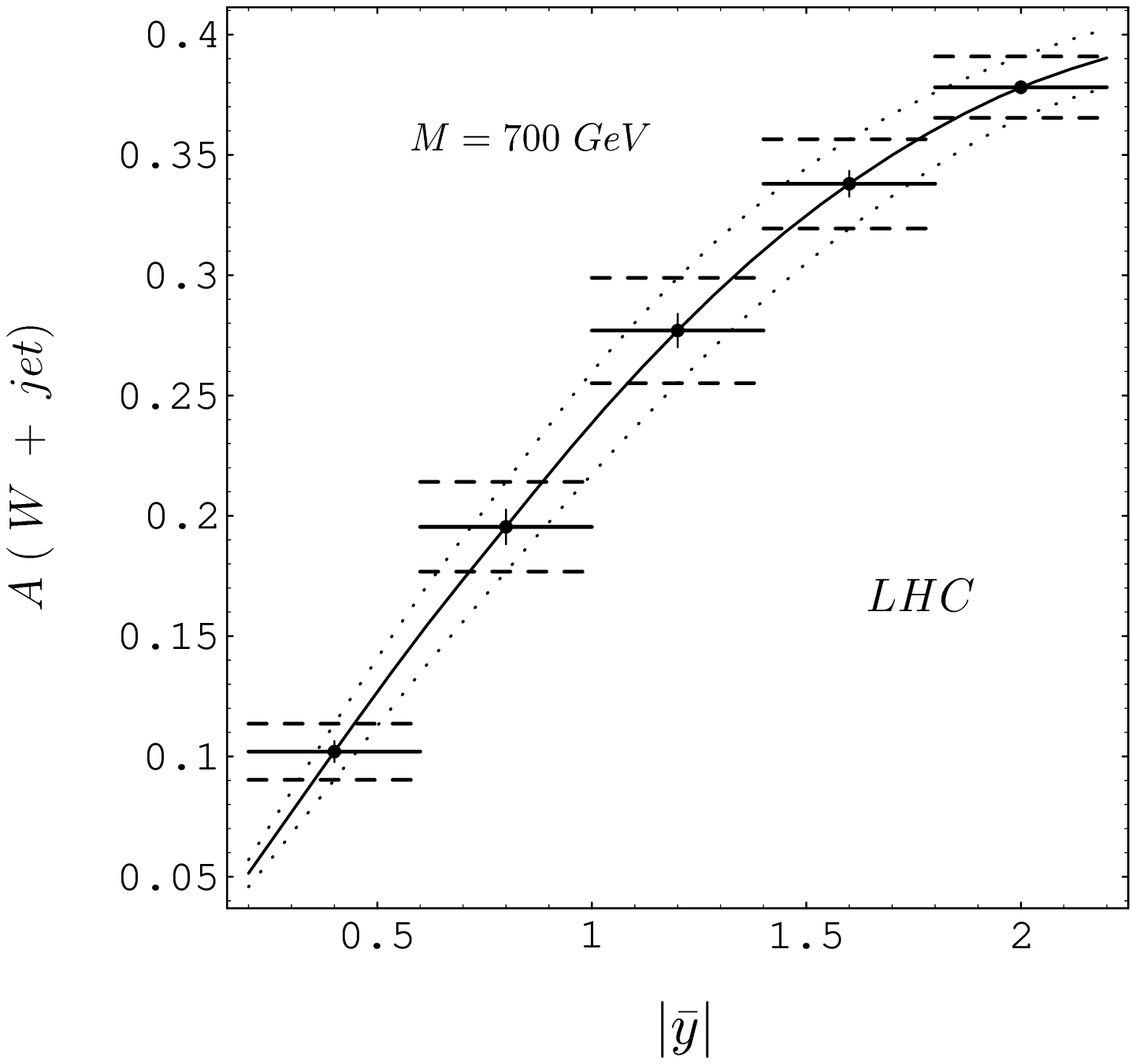,height=7.2cm}
\]
\vspace*{-3cm}\null\\
\hspace*{6.5cm} (a) \hspace{7.5cm}  (b)
\vspace*{2cm}
\[
\epsfig{file=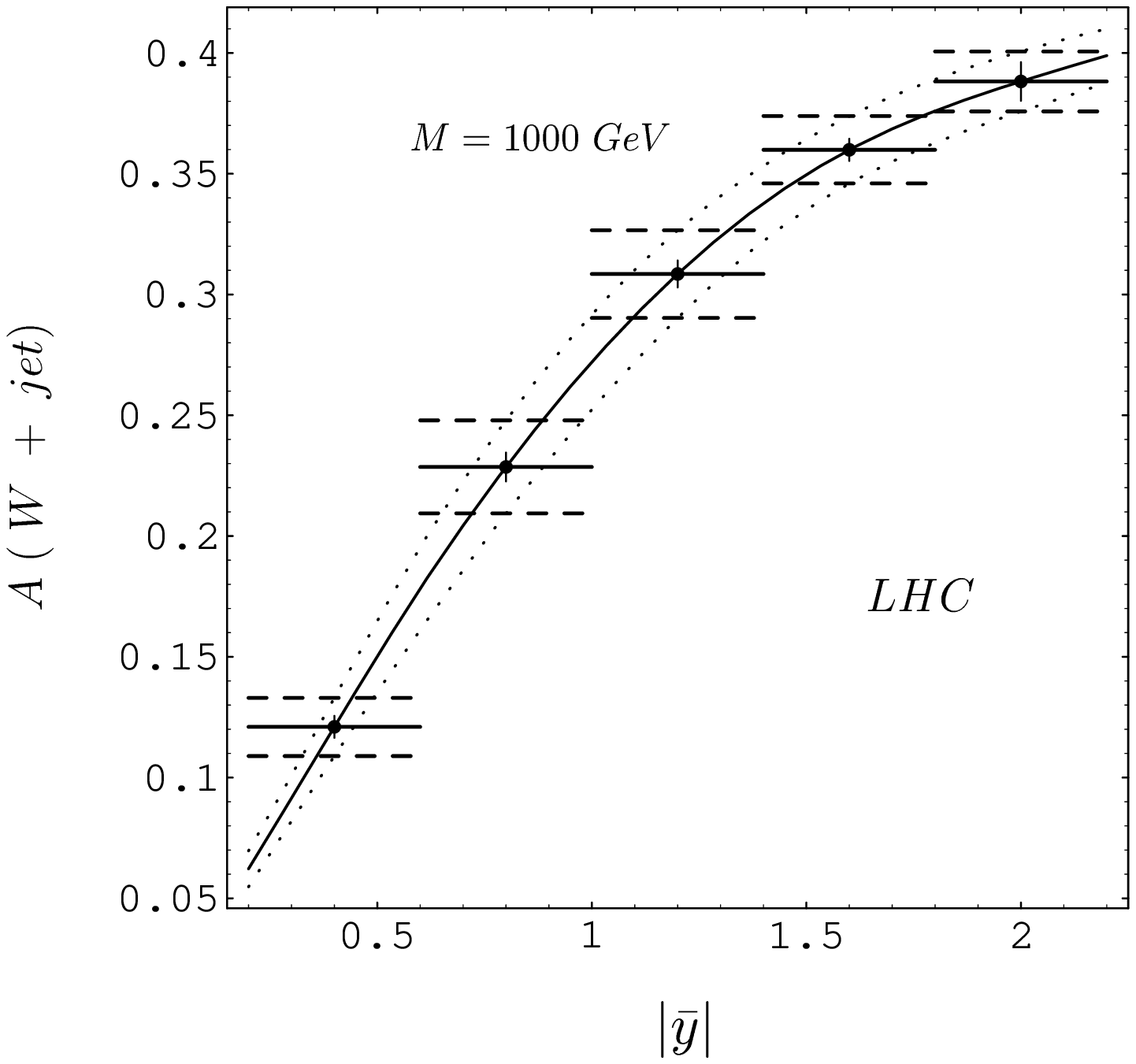,height=7.2cm}\hspace{0.5cm}
\epsfig{file=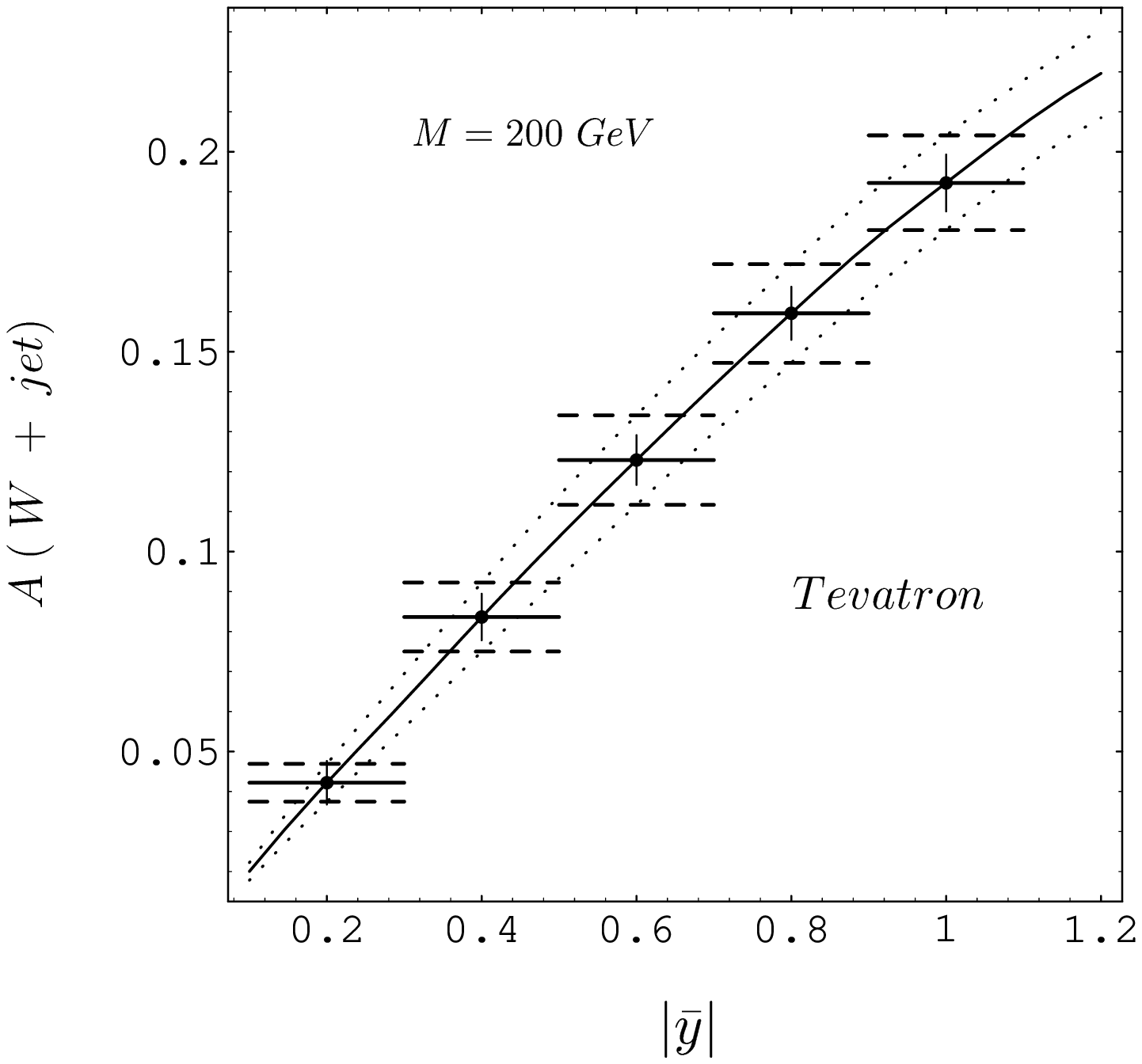,height=7.2cm}
\]
\vspace*{-3cm}\null\\
\hspace*{6.5cm} (c) \hspace{7.5cm}  (d)\\
\vspace*{1.cm}
\caption[1]{Asymmetry for $W$+jet production with invariant mass
$M$, as a function of the pair rapidity $\bar y$, for LHC
(a,b,c) and the upgraded Tevatron (d). The band between the two dotted
lines is due to the present uncertainty on the gluon distribution
function. The solid line is the center of the band. The error bar shows
the accuracy at which the asymmetry can be measured for each
indicated bin (dashed lines) taking into account statistical errors and
uncertainties on quark distribution functions.}
\label{Wprod}
\end{figure}

\begin{figure}[p]
\vspace*{-8cm}
\[
\epsfig{file=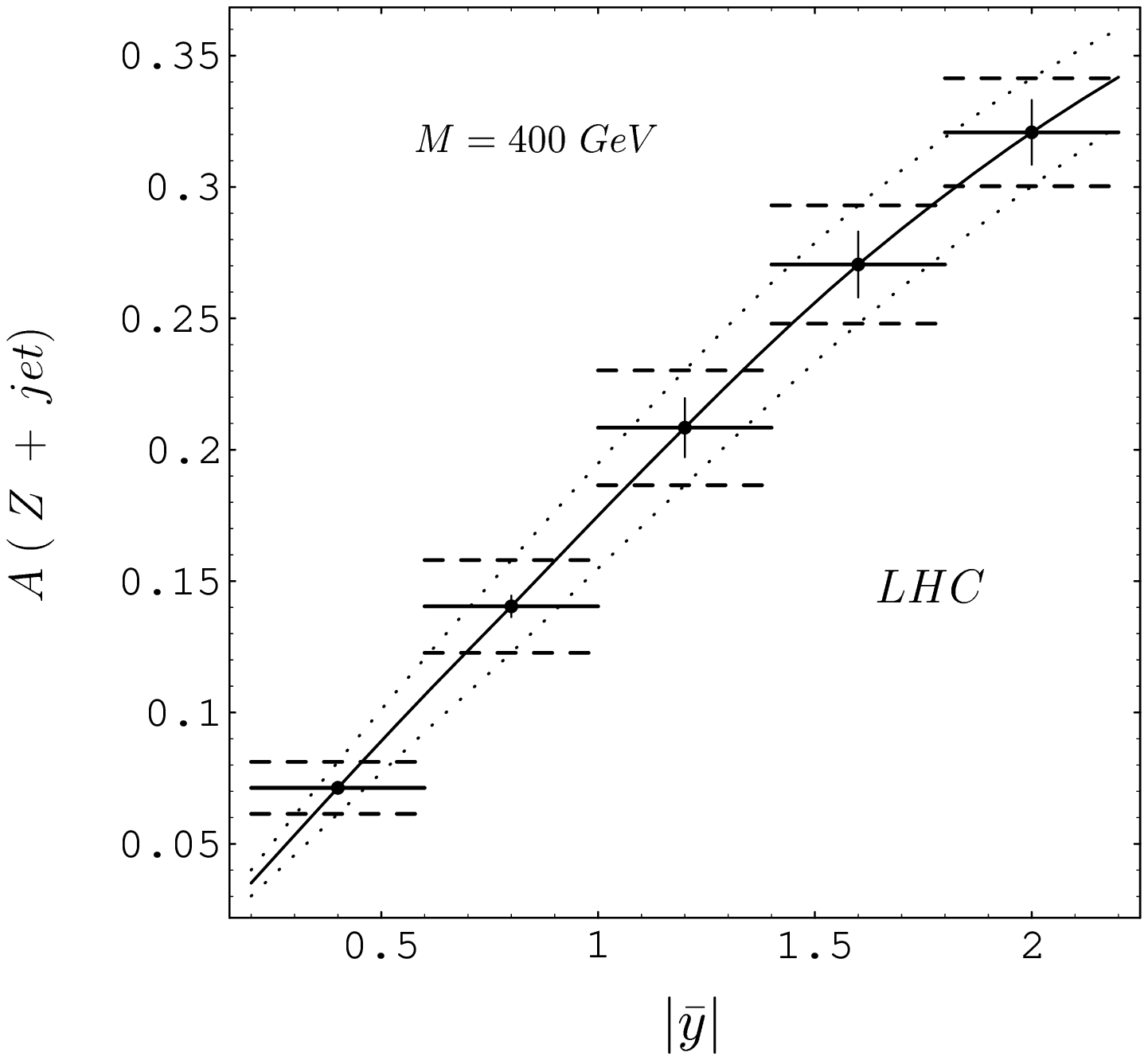,height=7.1cm}\hspace{0.4cm}
\epsfig{file=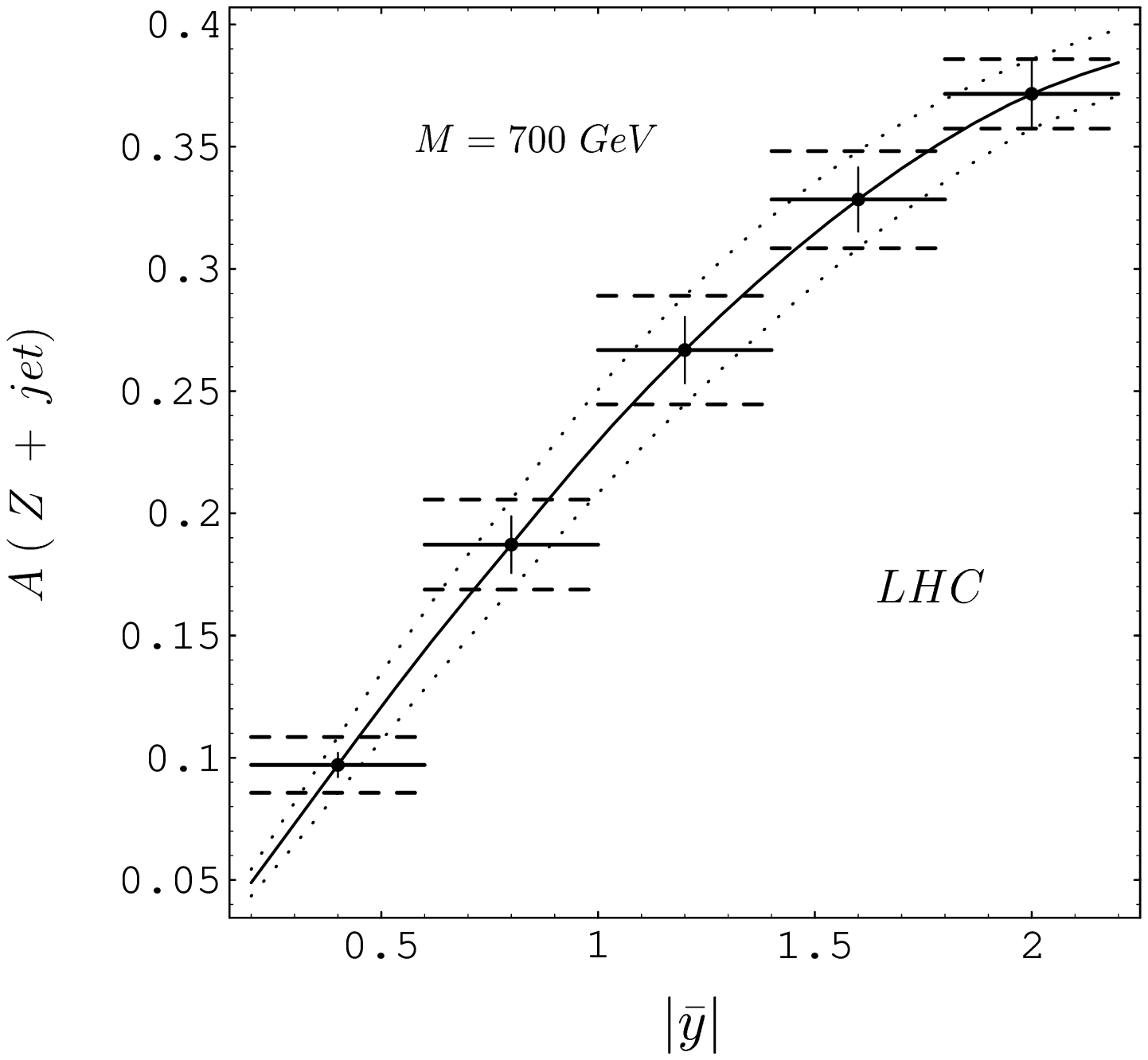,height=7.1cm}
\]
\vspace*{-3cm}\null\\
\hspace*{6.5cm} (a) \hspace{7.5cm}  (b)
\vspace*{2.cm}
\[
\epsfig{file=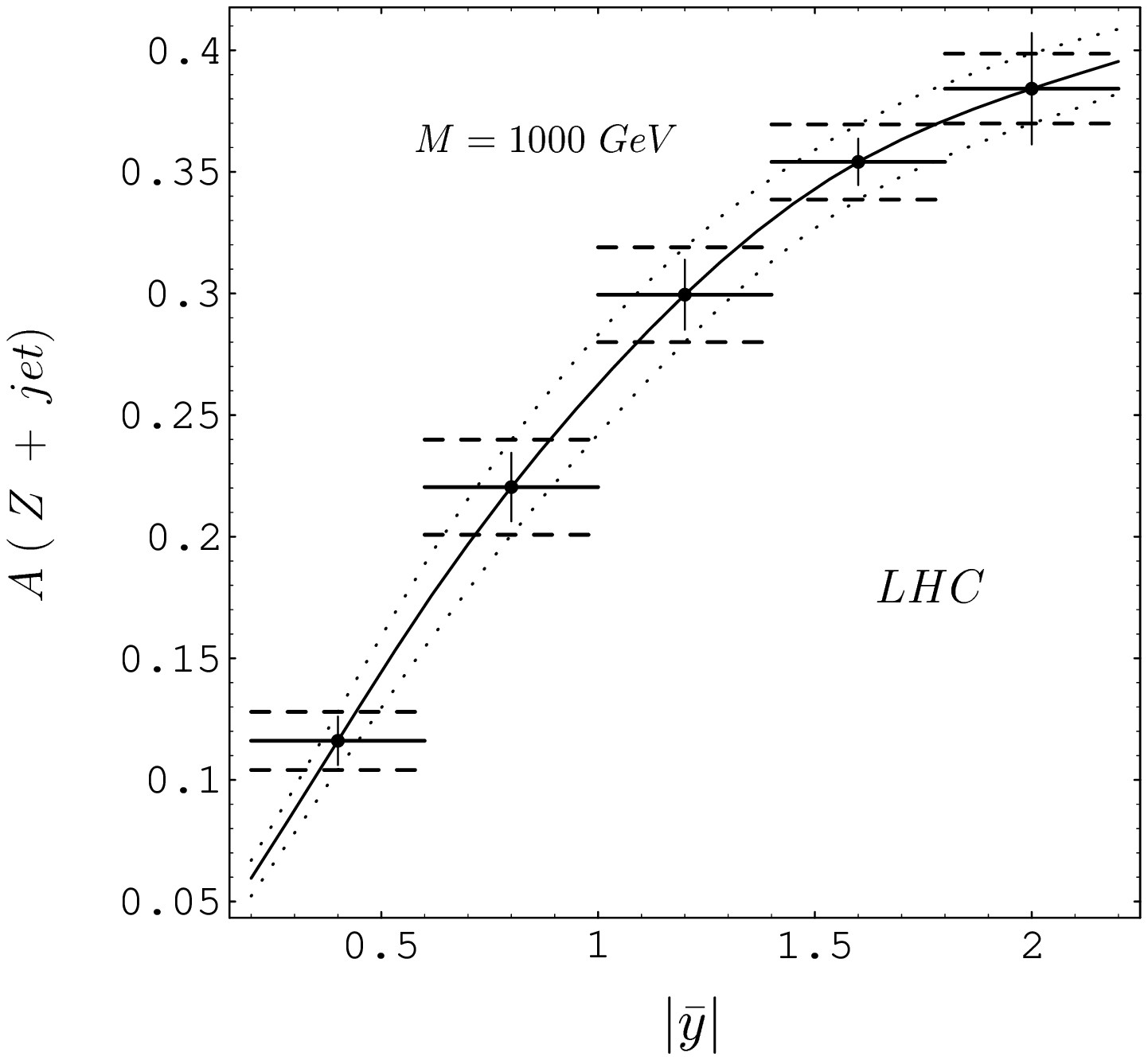,height=7.1cm}\hspace{0.4cm}
\epsfig{file=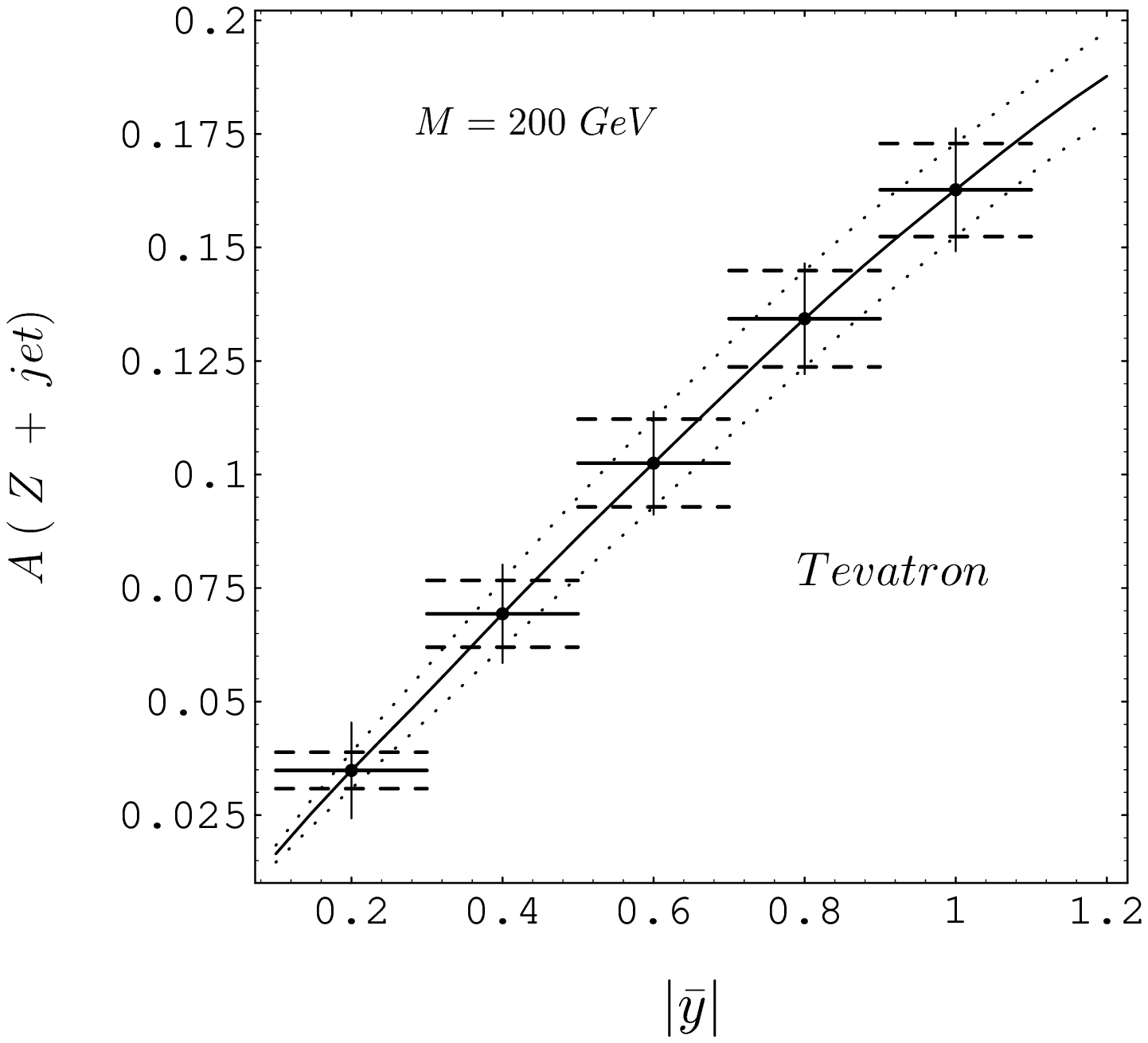,height=7.1cm}
\]
\vspace*{-3cm}\null\\
\hspace*{6.5cm} (c) \hspace{7.5cm}  (d)\\
\vspace*{1.cm}
\caption[1]{Asymmetry for $Z$+jet production with invariant mass
$M$, as a function of the pair rapidity $\bar y$, for LHC
(a,b,c) and the upgraded Tevatron (d). Same caption as in Fig.1.}
\label{Zprod}
\end{figure}

\begin{figure}[p]
\vspace*{-8cm}
\[
\epsfig{file=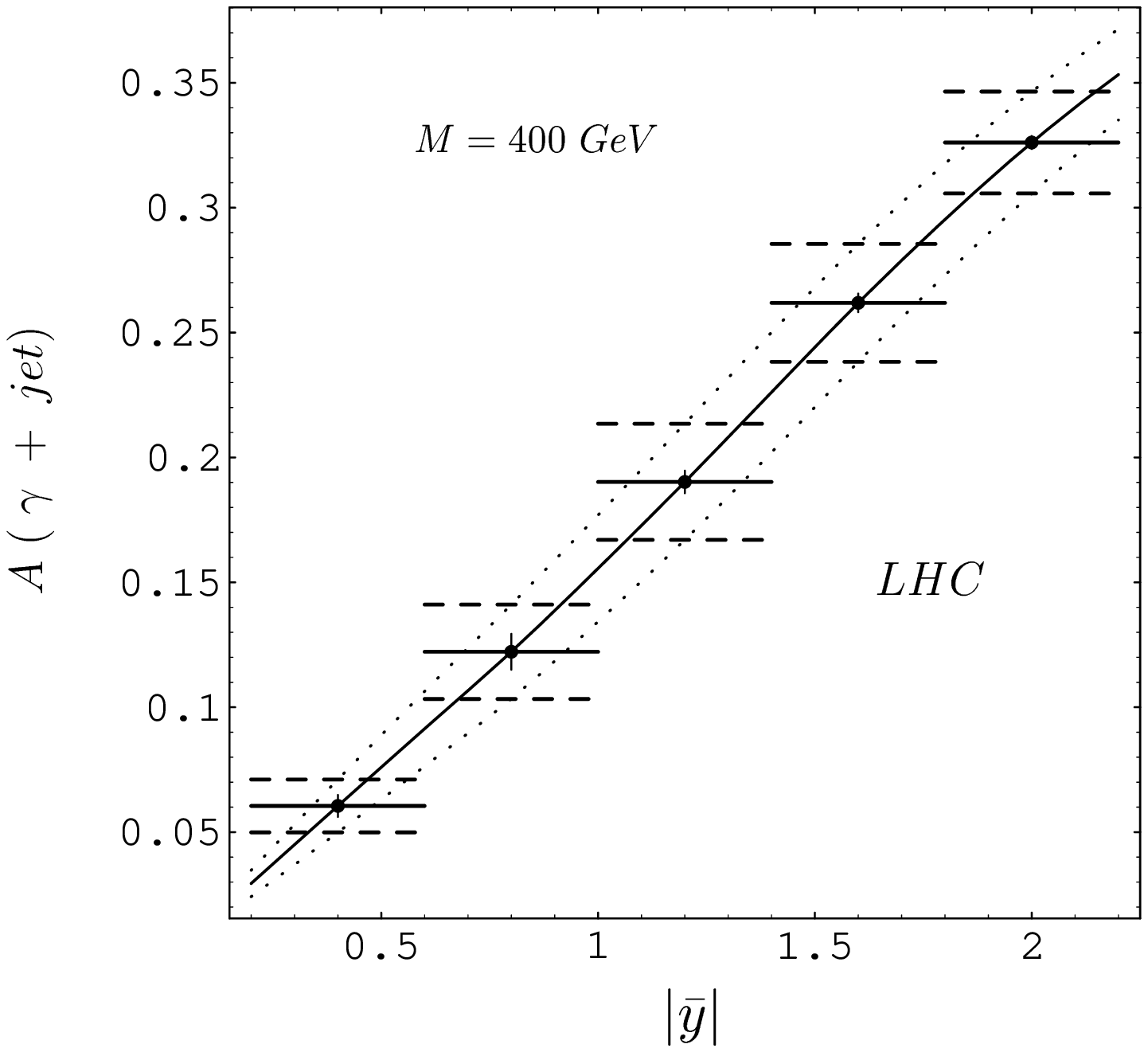,height=7.1cm}\hspace{0.4cm}
\epsfig{file=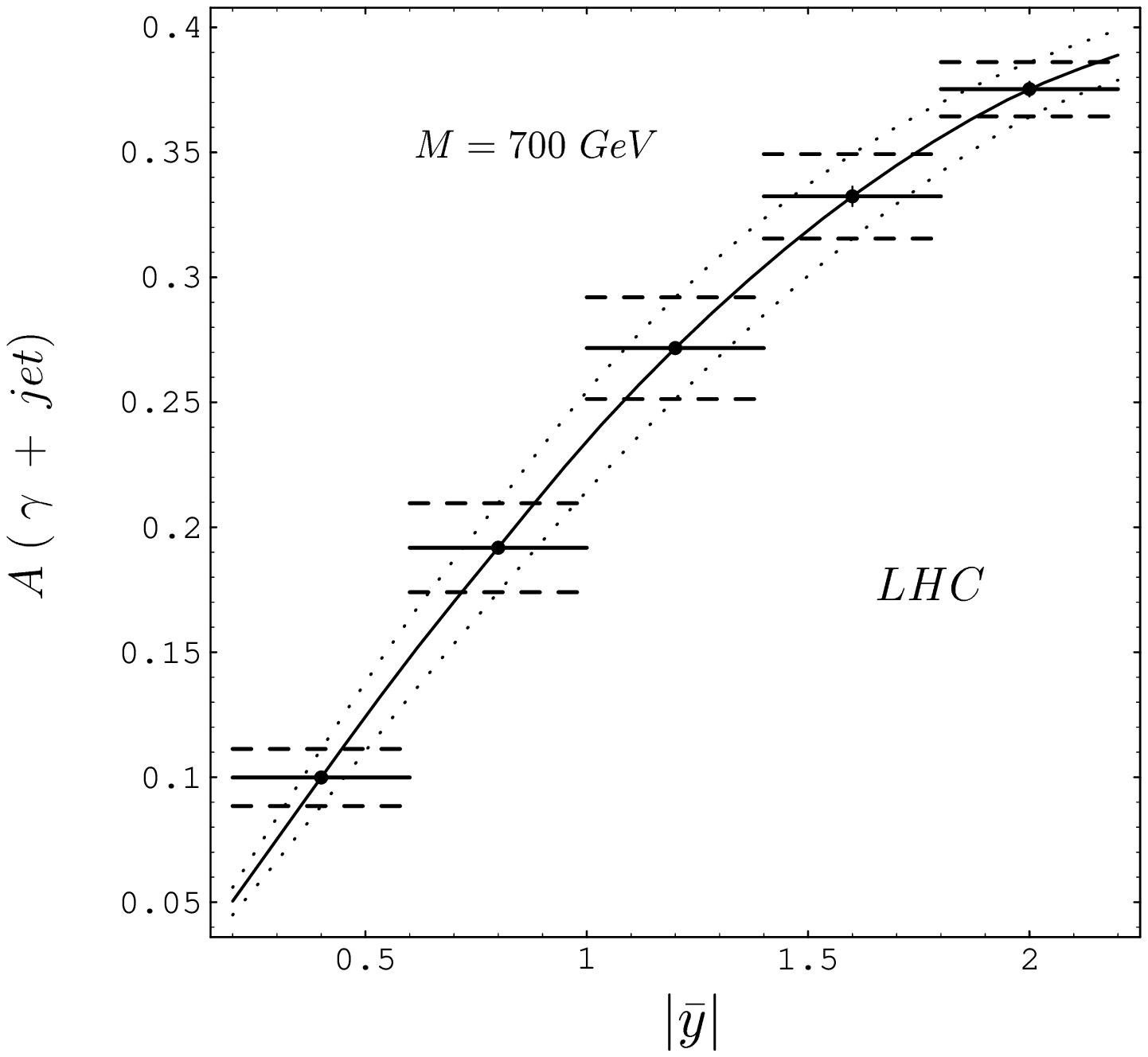,height=7.1cm}
\]
\vspace*{-3cm}\null\\
\hspace*{6.5cm} (a) \hspace{7.5cm}  (b)
\vspace*{2cm}
\[
\epsfig{file=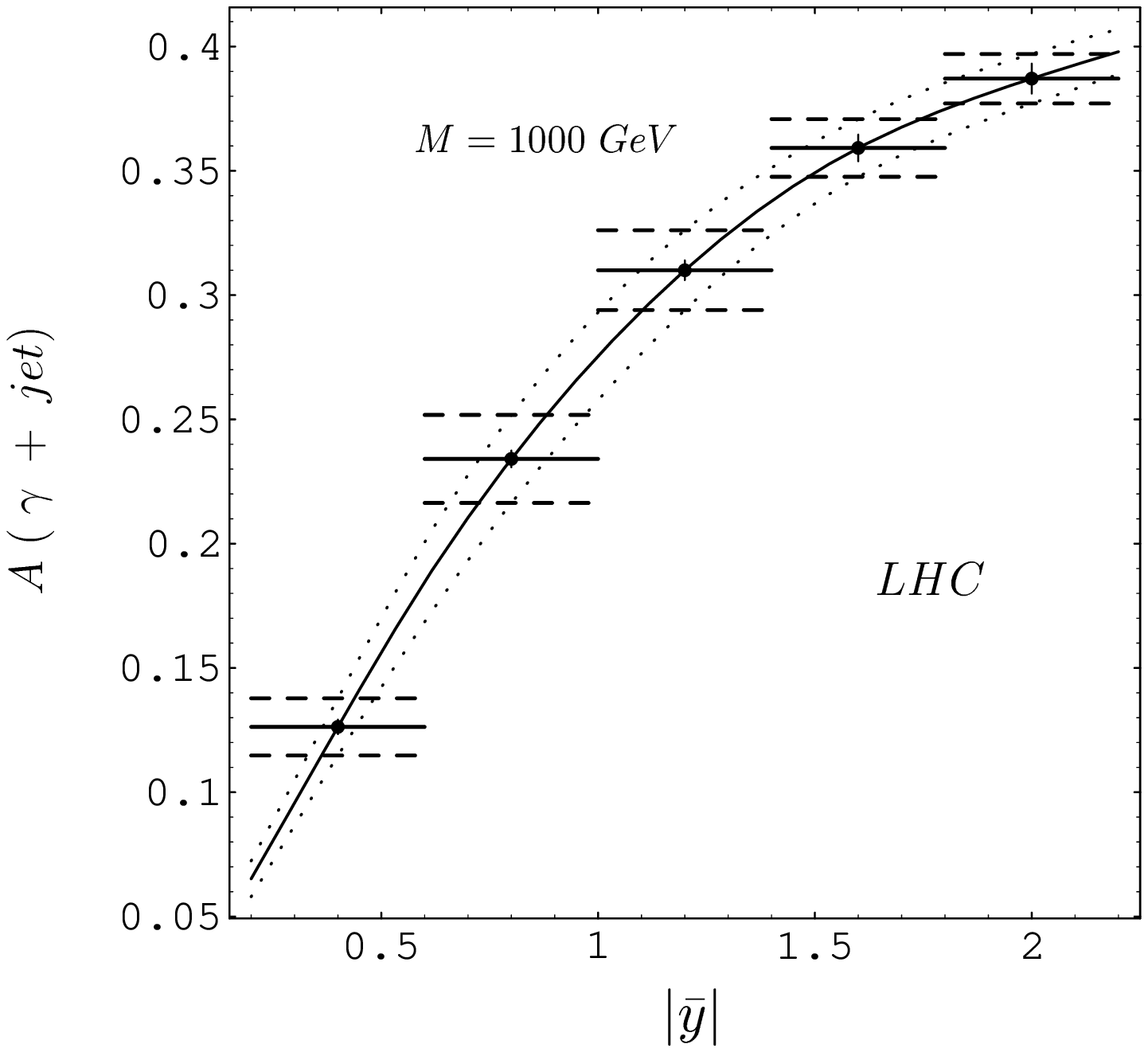,height=7.1cm}\hspace{0.4cm}
\epsfig{file=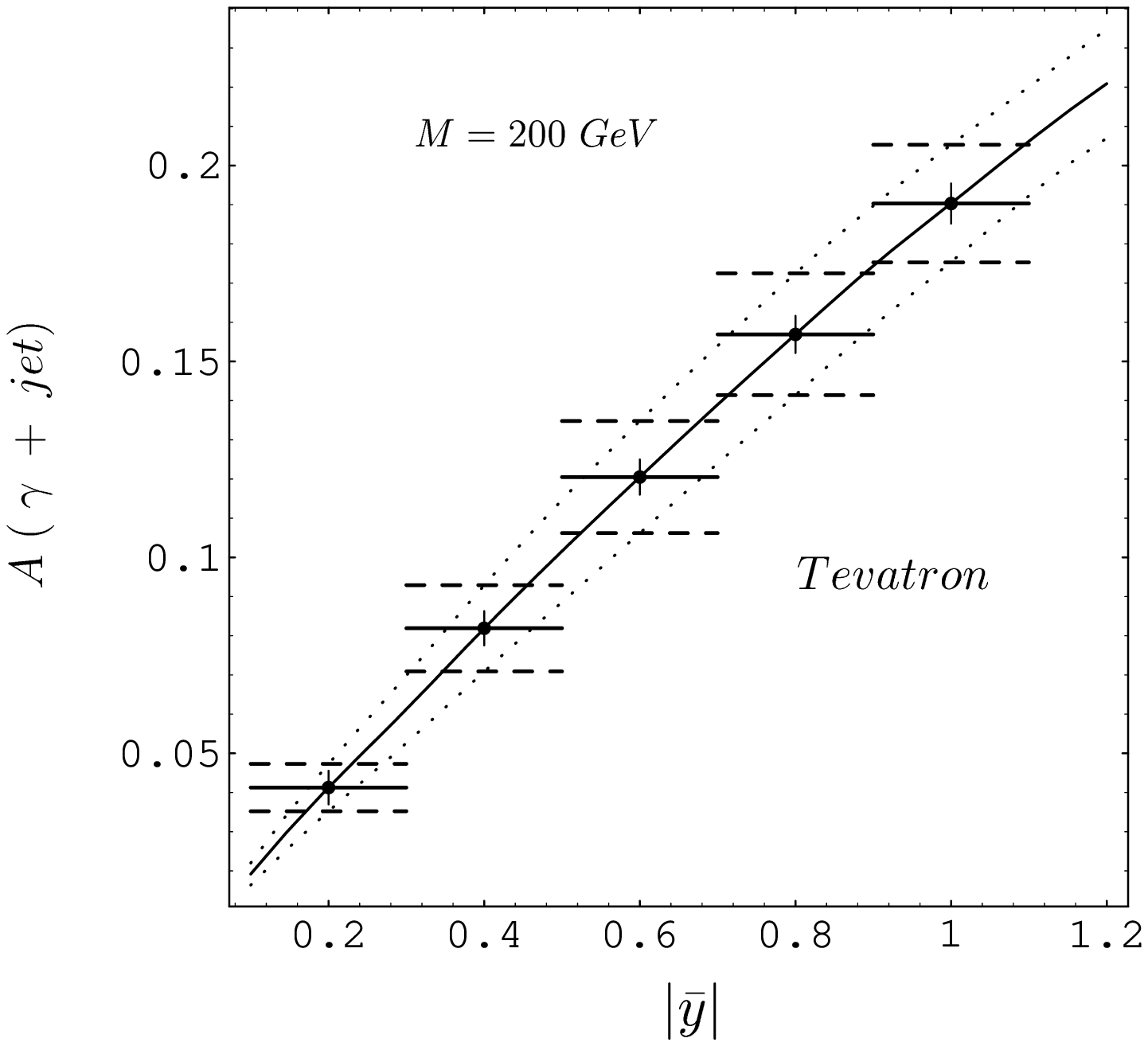,height=7.1cm}
\]
\vspace*{-3cm}\null\\
\hspace*{6.5cm} (c) \hspace{7.5cm}  (d)\\
\vspace*{1.cm}
\caption[1]{Asymmetry for $\gamma$+jet production with invariant mass
$M$, as a function of the pair rapidity $\bar y$, for LHC
(a,b,c) and the upgraded Tevatron (d). Same caption as in Fig.1.}
\label{Gammaprod}
\end{figure}

\begin{figure}[p]
\vspace*{-6cm}
\begin{center}
\mbox{
\epsfig{file=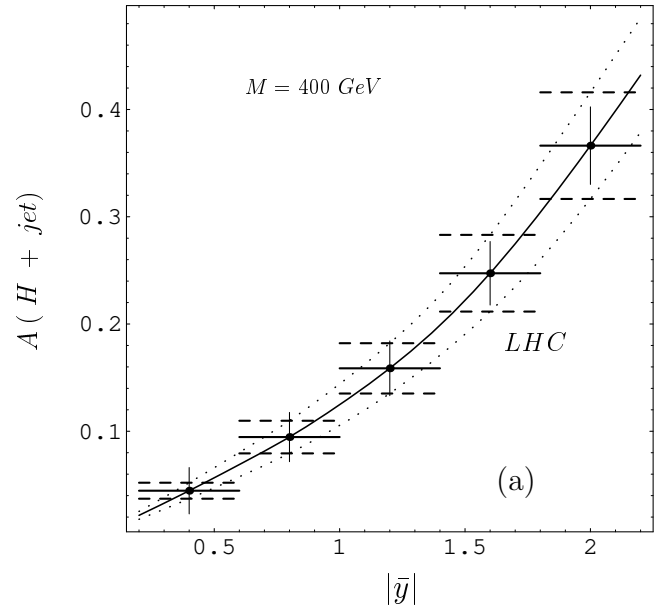,height=8cm}}
\vspace*{-2.5cm}
\end{center}
\hspace{10cm} (a)\\
\vspace*{1.5cm}
\begin{center}
\mbox{
\epsfig{file=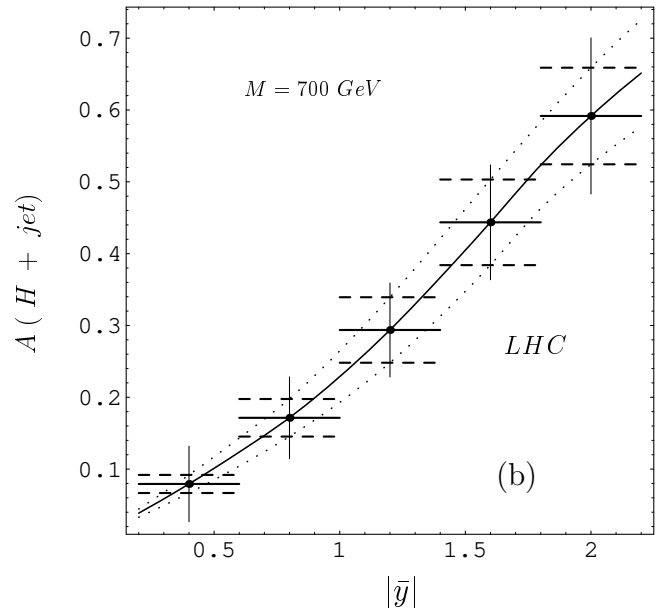,height=8cm}}
\vspace*{-2.5cm}
\end{center}
\hspace{10cm} (b)\\
\vspace*{2cm}
\caption[1]{Asymmetry for $H$+jet production with invariant mass
$M$, as a function of the pair rapidity $\bar y$, for LHC
(a,b). Same caption as in Fig.1.}
\label{Hprod}
\end{figure}

We first want to see what uncertainty our present ignorance of the
parton distributions in general, and of the gluon distribution
in particular, imply for the above asymmetry.
This is done by using the existing fits for the parton
distributions \cite{MRST,
CTEQ4M, Reya98}, which provide an estimate of the bands inside which the
asymmetry may lie, for various values of the $V$+jet invariant
mass $M=\sqrt{\s}$ and the rapidity $\bar y$ of the $V$+jet
pair. These bands are given by the dotted lines in
Figs.\ref{Wprod}, \ref{Zprod}, \ref{Gammaprod}, \ref{Hprod} for the
$W^\pm$, $Z$, $\gamma$ and $H$  production cases respectively,
while the full line determines the ``averaged''
asymmetry in each of these cases. In all figures we use
$\epsilon = 0.05$  to cut the forward or
backward $V$ production, with respect to the c.m.
angle; but the results are not sensitive to the exact magnitude
of this value; (compare (\ref{Asym1})). \par

In the figures  we also include the antenna pattern contribution
generated by a soft gluon jet ``close'' to the hard jet generated by the
parton $c$; compare (\ref{k}, \ref{R1R2}, \ref{kTmaxmin}).
This is integrated in the phase space region \cite{Khoze}, \cite{Khulmann}:
\bqa
0.7 \lsim \delta R \lsim 1 \ , &
70 GeV \lsim k_T \lsim 200 GeV  & \mbox{ for LHC} \, \\
0.7 \lsim \delta R \lsim 1.3 \ , &
10 GeV \lsim k_T \lsim 30 GeV  &
\mbox{ for Tevatron} ~(\sqrt s=2TeV) \ .
\eqa
The effect of such antenna contributions is to
\underline{increase} the
forward and backward cross sections and the asymmetry by roughly
 10\%, in all cases. \par

For the $\gamma$ production, we also  include the
fragmentation contribution from (\ref{sigmafrag}). It turns out
that for the strong photon isolation constrain imposed  by
$\max (E_{had})= 4GeV$ (compare (\ref{sigmafrag},
\ref{zmin}), the fragmentation contribution to the forward or
backward cross sections are at the 10\% level. This
photon fragmentation contribution tends to
\underline{decrease} the asymmetry by roughly 10\%.\par

Thus, the bands in Figs.1-4, give an estimate of the present
uncertainty on these asymmetries, which is due to our present
ignorance of the gluon distributions.
To see what can be achieved by LHC and/or the upgraded Tevatron,
we indicate on the same figures the uncertainties of
a possible (future) measurement of these asymmetries, assumed to
lie  on  the ``averaged''  solid line.
These uncertainties include the statistical ones  implied by
the integrated luminosities (2 experiments with 3 years of running)
of $600 fb^{-1}$ and $12 fb^{-1}$ for the LHC ($\sqrt s =14~TeV$)
and the upgraded Tevatron ($\sqrt s =2~TeV$) respectively.
For $W$ and $Z$,
we only retain leptonic decay modes with branching fractions
$Br=0.16$ and $0.09$, respectively. One should notice that the
$Z$ + jet final state is experimentally very clean.

For  $\gamma $ and $H$
we assume a detection efficiency of $0.8$ and $0.01$ respectively.
However  $\gamma $ + jet final state suffers from a background
from neutral pions, which will limit the accuracy of such a final state.
The detection efficiency in the Higgs case
is of course very much dependent on the value of the
Higgs mass which controls the value of the branching ratios in
$\gamma\gamma$, $WW^*$, $ZZ^*$ etc,
and the value of $0.01$ is only chosen for orientation. \par

In the illustrations given in Figs.1-4 we have computed
the statistical uncertainties taking bins of $\delta \bar y=\pm0.02$
(indicated by the horizontal lines in the figures) and
$\delta \tau = (\pm300~ GeV \cdot M)/s$.
Eq.(\ref{xaxb}) gives the relation between the
kinematical domains  ($\tau=M^2/ s,~ \bar y$) and  ($x_a,
~x_b$). Note that
reducing the size of the bins would make the domain in ($x_a,
~x_b$) more
restrictive, but it would
simultaneously increase the statistical error.\par

To these statistical
uncertainties we quadratically add  the uncertainties
implied by the present knowledge of the quark distribution
functions, estimated by their spread among the
various models. One should take into account that at LHC the
lepton pseudorapidities from weak gauge bosons decays will
provide the key to measure the $q$, $\bar q$ distribution
functions within an accuracy of $\simeq 1 \%$  \cite{DPZ}.
The resulting total uncertainties are indicated by the vertical
error bars at the central point of each bin.\par

 Figs.1-4 show the kinematical domains
where the measurement of the asymmetries can provide
useful constraints on $g(x)$. In order to get informations
on $g(x)$ at large $x$, one needs measurements
at large $M$ and $\bar y$; compare (\ref{xaxb}).
The number of events is however
decreasing with $M$ and $\bar y$. For LHC we see that asymmetries
for $W$ and $\gamma$ production
are  useful (i.e. at the percent level)
 for $M \lsim 1000~GeV$ and $|\bar y| \lsim 2$,
which means $x $ values up to $x \simeq 0.3$.
The results for $Z$ production
are given in Fig.\ref{Zprod}. The
expected sensitivity to $g(x)$ is weaker than in the $W,\gamma$
cases and it can be useful only up to $x \simeq 0.1$.
The Higgs case is shown in Fig.\ref{Hprod}. One cannot expect an
improvement on the determination of $g(x)$ from it, but it can
contribute to consistency checks, particularly with
respect to excluding possible New Physics contributions.\par

For the upgraded Tevatron the lower energy ($\sqrt s =2TeV $)
and the lower luminosity forces us to restrict to $M
\lsim 200 GeV$, for which we could reach useful information (at the
percent level) only
for $x \lsim 0.1$ and only from  $\gamma $ production (compare
Fig.\ref{Gammaprod}), the $W$ case giving only consistency checks.\\

\section{Conclusions}

In this paper we have
shown that forward-backward asymmetries with respect to
the subprocess c.m. scattering angle in
$V$+jet  production at hadron colliders,
are directly proportional to the
gluon distribution function $g(x)$. We have then studied how
measurements of these asymmetries can improve our knowledge of this
gluon distribution.\par

We have considered $V$+jet  production at the upgraded Tevatron and
at the LHC, for the cases $V=W^{\pm},Z,\gamma$ and $H$.
In order to make our treatment more realistic, we  have added to
the leading $V$+jet cross section, the
contribution from the soft gluon emission in the form of an antenna
pattern distribution. In the case of the photon we have also added
the photon fragmentation contribution arising from the various
subprocesses involving two body scattering among quarks
and gluons.\par

We have made an estimate of the accuracy at which
these asymmetries can be
measured at the upgraded Tevatron and
at the LHC in several bins of $V$+jet invariant mass and c.m.
rapidity. Each bin corresponds to a certain domain in the fraction of
momentum $x$ carried by the partons. We have given the
kinematical domains where this accuracy would allow a
substantial improvement in the
knowledge of the gluon distribution $g(x)$. More explicitly:
\begin{itemize}
\item
At the upgraded Tevatron ($\sqrt{s}=2~TeV$, with an integrated
luminosity of $12~fb^{-1}$) and using mainly the $\gamma$+jet process,
one can expect to reach an accuracy of the percent level up to about
$x\simeq0.1$.
\item
At LHC ($\sqrt{s}=14~TeV$, with an integrated
luminosity of $600~fb^{-1}$), mainly through the $\gamma$+jet and
$W$+jet processes (and at a weaker level the $Z$+jet one) one should
test $g(x)$ at less than one percent up to $x\simeq0.3$.
\item
A comparison of the results in the various modes, including
the $H$+jet one (which is by itself
not accurate enough to severely constrain
$g(x)$)
should give consistency checks and constrain possible non standard
contributions. In this respect, the radiation
antenna pattern may also help.
\end{itemize}

 The present study should just be taken as ``a stone to be used for the
construction of the building''. Obviously the
experimental information provided by these asymmetries should be
inserted in a global analysis
involving all other observables and all parton distributions.
In this context it should be remembered that the information
contained in these asymmetries arises
from subprocess-terms which are completely different from
those determining the present prompt
photon \cite{WA70,  E706} and the dijet production cross sections
at the Tevatron \cite{dijet}. The simple-minded analysis
presented here was just
aimed at showing the peculiar role of these asymmetries
for the purpose
of determining the gluon distribution function,
provided - as it will be the case in future hadronic
colliders - high statistics will allow to take into
account this forward backward asymmetry in a global
analysis.
\par

\newpage
\renewcommand{\theequation}{A.\arabic{equation}}
\setcounter{equation}{0}
\setcounter{section}{0}

{\large \bf Appendix: Parton expressions for
$V$ production at a $pp$ Collider.}

\vspace*{0.5cm}
For \underline{{\bf H}  production} the relevant
quark distribution to be used in (\ref{ppVjet}, \ref{ppVjet+g})
are
\bqa
\Sigma_H(x)&& = \Sigma(x) =\sum_q \left [ q(x, Q^2)+ \bar q(x, Q^2) \right
] \ \ , \label{SigH} \\
\wtil{\Sigma}_H(x_a, x_b) &&= \sum_q \left [ q(x_a, Q^2)
\bar q(x_b, Q^2)
+ \bar q(x_a, Q^2) q(x_b, Q^2)  \right ]  \ \ , \label{SigHtil}
\eqa
where the sum is over all light quark flavours.
The subprocess cross sections can be found in
\cite{HiggsLHC2, HiggsLHC1}. Note that the gluon-gluon term in
(\ref{ppVjet}) contributes only for  Higgs production,
and it should be ignored in the ($W, ~Z, ~\gamma $)
cases below. \par

\vspace*{0.5cm}
 For \underline{{\bf W} production} (${\bf W} \equiv W^++W^-$)
\bqa
\Sigma_W(x)& =  &\Sigma(x)  \ \ , \label{SigW}\\
\wtil{\Sigma}_W(x_a, x_b) &= & d(x_a, Q^2)\bar u(x_b, Q^2)
+\bar u(x_a, Q^2) d(x_b, Q^2)
+ \bar d(x_a, Q^2) u(x_b, Q^2) \nonumber \\
& + & u(x_a, Q^2) \bar d(x_b, Q^2)
+  [ u \to c ~~,~~ d \to s]  \ , \ \  \label{SigWtil}
\eqa
provided we neglect the CKM matrix elements $V_{td}, ~ V_{ts}$.
The corresponding subprocess cross sections, to leading order in
QCD, are
\bqa
\frac{d \hat \sigma ( q \bar {q}^\prime \to W g)}{d \hat t}  &=&
\frac{\alpha_s \sqrt{2} G_F \mwd}{4 \hat s^2} ~~ \frac{8}{9}
~~ \frac{ \t^2+ \u^2 +2 \mwd \s}{\t \u} \ , \label{qqWg} \\
\frac{d \hat \sigma ( g q  \to W q^\prime )}{d \hat t}  &=&
\frac{\alpha_s \sqrt{2} G_F \mwd}{4 \hat s^2} ~~ \frac{1}{3}
~~ \frac{ \s^2+ \u^2 +2 \mwd \t}{-\s \u} \ , \label{gqWq} \\
\frac{d \hat \sigma ( q g   \to W q^\prime )}{d \hat t}  &=&
\frac{\alpha_s \sqrt{2} G_F \mwd}{4 \hat s^2} ~~ \frac{1}{3}
~~ \frac{ \s^2+ \t^2 +2 \mwd \u}{-\s \t} \ . \label{qgWq}
\eqa \par

\vspace*{0.5cm}
Correspondingly,  for  \underline{{\bf Z } production}
\bqa
\Sigma_Z(x)& = & \sum_q (g_{Lq}^2+g_{Rq}^2)
\left [ q(x, Q^2)+\bar q(x, Q^2) \right
] \ , \label{SigZ} \\
\wtil{\Sigma}_Z(x_a, x_b) &= & \sum_q  (g_{Lq}^2+g_{Rq}^2)
\left [ q(x_a, Q^2) \bar q(x_b, Q^2)
+ \bar q(x_a, Q^2)  q(x_b, Q^2) \right ]  \ \ \label{SigZtil}  ,
\eqa
where the Z-couplings to the $L$ and
$R$ quarks $g_{Lq}=t^{(3)}_q-e_q \swd$ and $g_{Rq}= -e_q \swd$,
are absorbed
to the definition of the $\Sigma_Z, ~\wtil{\Sigma}_Z$
distributions. In this normalization,
the corresponding subprocess cross sections are given by
(\ref{qqWg}-\ref{qgWq}), after replacing $\mw \to \mz$.\par

\vspace*{0.5cm}
 Finally for \underline{{\boldmath $ \gamma $ } production}
we have
\bqa
\Sigma_\gamma (x)& = & \sum_q e_q^2
\left [ q(x, Q^2)+\bar q(x, Q^2) \right
] \ , \label{SigGamma} \\
\wtil{\Sigma}_\gamma(x_a, x_b) &= & \sum_q  e_q^2
\left [ q(x_a, Q^2) \bar q(x_b, Q^2)
+ \bar q(x_a, Q^2)  q(x_b, Q^2) \right ]  \ \ \label{SigGammatil}.
\eqa
Since  the square of the quark charge  $e_q^2$
has been inserted in the definition
(\ref{SigGamma}, \ref{SigGammatil}), it is removed from
the subprocess cross sections, which is thus written  to leading QCD
order as
\bqa
\frac{d \hat \sigma ( q \bar q  \to \gamma  g)}{d \hat t}  &=&
\frac{\pi \alpha_s \alpha }{\s^2} ~~ \frac{8}{9}
~~ \frac{ \t^2+ \u^2 }{\t \u} \ , \label{qqGammag}\\
\frac{d \hat \sigma ( g q  \to \gamma  q )}{d \hat t}  &=&
\frac{\pi \alpha_s \alpha }{ \hat s^2} ~~ \frac{1}{3}
~~ \frac{ \s^2+ \u^2 }{-\s \u} \ , \label{gqGammaq}\\
\frac{d \hat \sigma ( q g   \to \gamma  q )}{d \hat t}  &=&
\frac{\pi \alpha_s \alpha }{ \hat s^2} ~~ \frac{1}{3}
~~ \frac{ \s^2+ \t^2 }{-\s \t} \ . \label{qgGammaq}
\eqa \par


\clearpage
\newpage

\clearpage
\newpage

\clearpage
\newpage

\clearpage
\newpage

\end{document}